# Symptoms of complexity in a tourism system

**Rodolfo Baggio**

Master in Economics and Tourism, Bocconi University, Milan, Italy
School of Tourism, The University of Queensland, Australia

rodolfo.baggio@unibocconi.it

## Abstract

Tourism destinations behave as dynamic evolving complex systems, encompassing numerous factors and activities which are interdependent and whose relationships might be highly nonlinear. Traditional research in this field has looked after a *linear* approach: variables and relationships are monitored in order to forecast future outcomes with simplified models and to derive implications for management organisations. The limitations of this approach have become apparent in many cases, and several authors claim for a new and different attitude.

While complex systems ideas are amongst the most promising interdisciplinary research themes emerged in the last few decades, very little has been done so far in the field of tourism. This paper presents a brief overview of the complexity framework as a means to understand structures, characteristics, relationships, and explores the implications and contributions of the complexity literature on tourism systems. The objective is to allow the reader to gain a deeper appreciation of this point of view.

## 1   Introduction

Tourism is one of the most important economic activities in the World. The revenue generated has become a very important resource and a key factor in the balance of payment for many countries and regions and has been a major contributor to their economic growth. As a natural consequence, it has become, in the last decades, a discipline studied by a growing number of researchers, practitioners, experts and consultants. Their main objective is to describe and to understand the composition and the dynamics of the sector and, based on this knowledge, to be able to foresee future behaviours of the system's components. This is the basis for a great number of decisions, involving single operators as well as governing bodies at different levels (Hall et al., 2004).

The task is not easy, tourism is difficult to measure and analyse. The main reason resides in the fact that it is an "industry" with no traditional production functions, no consistently measurable outputs and no common structure or organisation across countries or even within the same country (see for example: OECD, 2000). Moreover, tourism activities traverse a number of traditional economic sectors and are generally not considered, as a whole, in national accounts.

The World Tourism Organisation's definition of tourism as comprising (UNWTO, 1995):
> the activities of persons travelling to and staying in places outside their usual environment for not more than one consecutive year for leisure, business and other purposes.

looks fuzzy if examined with the glasses of a scientist.

Too many different elements and interpretations fall into the terms contained in the official definition of tourism. The best proof is that all the official bodies responsible for "measuring" this phenomenon face a real challenge when it comes to identifying the "units" to be accounted for. Moreover, this definition poses a great challenge to all those seeking to model the phenomenon in order to foresee its behaviour.

The forecasting methodology of tourism demand has created numerous proposals over the past decades. Classical regressions, time series analysis, econometric models, qualitative methods and, more recently, neural network techniques, have been extensively explored and have generated innumerable attempts (Song & Witt, 2000; Witt & Witt, 2000). Nonetheless, the general accuracy of these models cannot be regarded as excellent and the countless specific limitations put to them are a good testimony of this.

Failures in economic forecasting are common since Thomas Malthus predicted in 1798 that population growth would overtake food production resulting in mass starvation in Great Britain. For the tourism sector this failure of forecasting may be particularly serious, given its fundamental economic importance for many regions across the world. The great majority of the models have at the basis the idea of a simplified, linearised, version of the tourism system and therefore they must comprise a fair amount of limitations, boundaries and restrictions.

This condition is common in a number of disciplines such as physics, biology, ecology, medicine, sociology, economics.

In recent years a new approach has emerged. Many scholars believe this approach is able to overcome the difficulties of describing "complex" systems and to give better representations and better tools to handle the issues involved. This approach starts with the realisation that the *reductionist* hypothesis born with the origin of "modern science" is limiting too much of our ability to describe the real world. The methods devised by Galileo, Newton, Laplace, and many others, can only give us a very limited power and, more importantly, are not able to return reasonable explanations for a wide number of phenomena.

In his seminal paper "More is different", Phil Anderson states (1972: 393):
> The workings of our mind and bodies, and of all the animate or inanimate matter of which we have any detailed knowledge, are assumed to be controlled by the same set of fundamental laws, which except under certain extreme conditions we feel we know pretty well" but "the ability to reduce everything to simple fundamental laws does not imply the ability to start from those laws and reconstruct the universe" and "at each stage entirely new laws, concepts, and generalizations are necessary, requiring inspiration and creativity to just as great a degree as in the previous one.

These ideas have contributed to set a new perspective in our view of natural phenomena, a new view which today is known as "science of complexity" (Waldrop, 1992). While complex systems ideas are amongst the most promising interdisciplinary research themes to have emerged in the last few decades, not much has been done, so far, in the field of tourism studies.

## 1.1 Chaos, complexity and tourism

Tourism has become an extremely dynamic system. Economic globalisation, fast changing customer behaviour, development of transportation and information technologies, modifications in the forms of organisations and labour, all strongly influence the industry. In this scenario the intensified marketing efforts of all tourism organisations have led to a more effective approach: the destination management approach (Ritchie & Crouch, 2003).

Many definitions have been proposed for the notion of tourism destination (TD), all rather problematic for the many and diverse aspects that are usually comprised in the concept (Framke, 2002). Here we characterise a TD as a geographical location (city, region, resort etc.), with a pattern of attractions, facilities and services, which tourists choose for a visit. In many places, a destination management organisation (DMO) is responsible for coordinating the resources and the companies operating at the location. The main typical responsibilities of a DMO are: policy enforcement, strategic and operational planning, marketing and developing the product offering (coordinating both public and private assets).

From a structural point of view, a destination can be seen as a system composed by a number (usually not too small) of elements which share some kind of relationship. The system evolves by responding to external and internal inputs. It may well be considered a complex adaptive system. A tourism destination comprises a number of elements: the tourism operators, the support structures, public and private organisations and associations. All of these elements have some kind of relationship among themselves and the possible nonlinearities in these relationships are well known and have been described several times (Farrell & Twining-Ward, 2004; Faulkner & Russell, 1997). Moreover, we can include in the system also elements not traditionally thought as belonging strictly to the tourism sector, but whose importance and role in this framework is undoubtedly very high.

An important, although rather scarce, strand of literature has pointed out the necessity to change attitude when studying tourism and tourism systems. In a pioneering work, Faulkner and Valerio (1995) start from the realisation of the deficiencies and the unreliability of many demand prediction and forecasting methods to call for the need of alternative ways to explain tourism phenomena. They propose the adoption of a chaos and complexity framework. The reductionist paradigm used in dividing a tourism system into some components, assuming that the relationships between them are stable and static is challenged as unable to provide meaningful elucidations of many outcomes (Faulkner & Russell, 1997; Russell, 2005, 2006)

Examples of how the chaos and complexity framework is able to provide meaningful explanations of the dynamical behaviour of a tourism system can be traced in a number of subsequent works. These discuss issues such as the dynamic and serendipitous development of destinations (Faulkner, 2000, 2002; Russell & Faulkner, 1998), the effects of unexpected events such as crises and natural disasters (Faulkner & Russell, 2001; Faulkner & Vikulov, 2001; Scott & Laws, 2005), the actions of entrepreneurs and their influence on the development of a tourism destination (Russell, 2006; Russell & Faulkner, 1999, 2004).

Some other researchers have followed this approach. McKercher (1999), for example, encourages the use of a chaos model for tourism. Tourism operates in a non-linear manner and the explanation may come, according to the author, by taking into account the complex interactions of the system's elements, combined with the influence of a large set of external factors. The value of the chaos and complexity framework in understanding the development of a destination and the role of small tourism business networks has also been discussed by Tinsley and Lynch (2001).

The main objective of this paper is to give a brief overview of the complexity framework and to explore the implications and contributions that the study of complex systems can give to the understanding of the tourism destination model. Continuing the line of research presented above, this work aims at complementing and reinforcing it by providing some quantitative evidence in support of this approach. This, it is hoped, will allow the reader to gain a deeper appreciation of this point of view.

The remainder of the paper is organised as follows. The next section gives an overview of the so-called "science of complexity" and the tools used to study a complex system; where possible, examples from the tourism area of studies will be given. Section 3 discusses tourism destinations in this framework giving some quantitative evaluations of the symptoms of complex behaviours in tourism systems. Section 4 discusses some implications of adopting the "complex system" approach in considering a tourism destination. The last part contains some conclusive remarks and indications for further research.

## 2  Complex systems

*Complexity* is the study of the structures which depend only in part on the nature of their constituents and whose global behaviours and characteristics cannot be deduced from the knowledge of their elementary building blocks. Complexity concerns the unforeseen adaptive capacities and the emergence of new properties in systems that arise as the quantity and the quality of the connections among individuals and organisations increase. The notion of complexity has numerous meanings in everyday speech. These are usually related to the size and the number of a system's components. From a "technical" point of view, we do not have universally accepted definitions, nor rigorous theoretical formalisations, of complexity. A rather intuitive depiction of a complex system may be given as: "*a system for which it is difficult, if not impossible to reduce the number of parameters or characterising variables without losing its essential global functional properties*" (Pavard & Dugdale, 2000).

A system is considered complex if its parts interact in a nonlinear manner. Simple cause and effect relationships among the elements rarely exist and instead a very little stimulus may cause unpredictably large effects or no effect at all. This nonlinearity of the interactions among the components is a major originator of a series of properties that are frequently used to characterise the complex behaviour of a system. It must be noted that, despite what happens with the natural language, there is a difference between the concepts of *complicated* and *complex*. In comparison to a complex system, a complicated one is a collection of an often high number of elements whose collective action is the cumulative sum of the individual ones. It can be decomposed in sub-elements and understood by analysing each one of them. Any modern machine (a computer, a car, an airplane, a satellite) comprises thousands, or even millions, of single pieces bound together. But, no matter how difficult it is, it is always possible to break up a *complicated* system into separate entities and study them individually, being sure that the final object will be the (linear) composition of them (Ottino, 2004; Procaccia, 1988).

On the contrary, a *complex* system can be understood only by considering it as a whole, almost independently by the number of parts composing it. A "simple" object made of only two elements, a double pendulum, a pendulum hanging from another pendulum, is well known to any physics student for its totally unpredictable, chaotic behaviour (under the basic Newtonian laws of motion). A "simple" school of fish, made up of a few dozen members, is able to adapt its behaviour to the external conditions without apparent organisation but following a very limited number of simple rules regarding local interaction, spacing and velocity (Reynolds, 1987).

A special class of complex systems is the one whose structure and components influence and are influenced by the external environment and in which the interactions among the elements are of a dynamic nature. In a *complex adaptive system* (CAS), the parts (Stacey, 1996: 10):

> interact with each other according to sets of rules that require them to examine and respond to each other's behaviour in order to improve their behaviour and thus the behaviour of the system they comprise.

Even in absence of a "formal theoretical definition" it is possible to identify a number of characteristics that may allow us to recognise a complex adaptive system (see for example: Cilliers, 1998; Stacey, 1996; Waldrop, 1992).

These main features can be summarised as follows:

- *non-determinism*. It is impossible to anticipate precisely the behaviour of a CAS even knowing the function of its elements. The dependence of the behaviour from the initial conditions is extremely sensitive and appears to be extremely erratic; the only predictions that can be made are probabilistic;

- *presence of feedback cycles (positive or negative).* The relationships among the elements become more important than their own specific characteristics and the feedback cycles can influence the overall behaviour of the system;

- *distributed nature.* Many properties and functions cannot be precisely localised, in many cases there are redundancies and overlaps; it is a distributed system;

- *emergence and self-organisation.* a number of emergent properties are not directly accessible (identifiable or foreseeable) from an understanding of the components. In a CAS, global structures may emerge when certain parameters go beyond a critical threshold. In these cases, generally, a new hierarchical level appears that reduces the complexity. The system evolves, increasing its complexity up to the next self-organisation process. One effect of such a characteristic is the capability to show a good degree of robustness to external (or internal) shocks. The system is capable to absorb the shock and to remain in a given state or regain the state unpredictably fast. At the critical points of instability the system will reorganise through feedback mechanisms. At a global level the system is homogeneous or symmetric; after a self-organisation process, however, symmetry is lost (breaks), one configuration dominates all others. From an empirical point of view it is virtually impossible to determinate why the system prefers one specific configuration instead of possible alternatives;

- *self-similarity.* It implies that the system considered will look like itself on a different scale, if magnified or made smaller in a suitable way. The self-similarity is evidence of a possible internal complex dynamics of a system. A CAS is at a critical state between a chaotic state and a completely ordered one, a condition that has been also called a self-organised criticality. If parameters $N$ and $z$, describe a self-similar system, they are related by a power-law relationship: $N \sim z^k$. A power law means that there is no "normal" or "typical" event, and that there is no qualitative difference between large and small fluctuations.

- *limited decomposability.* It is quite impossible, to study the properties of a dynamic structure by decomposing it into functionally stable parts. Its permanent interaction with the environment and its properties of self-organisation allow it to functionally restructure itself; only a "whole system" approach can explain CAS characteristics and behaviours.

In short, following Cilliers (1998), it is possible to characterise a system as complex and adaptive by listing these main properties:

- a large number of elements form the system;
- interactions among the elements are nonlinear and usually have a somewhat short range;
- there are loops in the interactions;
- complex systems are usually open and their state is far from equilibrium;
- complex systems have a history, the "future" behaviour depends on the past one
- each element is unaware of the behaviour of the system as a whole, it reacts only to information that is available to it locally.

Examples of complex adaptive systems include many real world ensembles: the patterns of birds in flight or the interactions of various life forms in an ecosystem, the behaviour of consumers in a retail environment, people and groups in a community, the economy, the stock-market, the weather, earthquakes, traffic jams, the immune system, river networks, zebra stripes, sea-shell patterns, and many others. Complexity also applies well to the world of economics. As Saari (1995: 222) writes, *"even the simple models from introductory economics can exhibit dynamical behaviour far more complex than anything found in classical physics or biology."* Many features of an economy present difficulty for the "linear" mathematics usually employed.

The tourism sector, as an economic activity, shares many of these characteristics. A destination comprises many different companies and organisations. The relationships among them exhibit a wide diversity and have been described in many different ways (Buhalis, 2000; Michael, 2003; Pavlovich, 2003; Pavlovich & Kearins, 2004; Smith, 1988), but, very often, they do not have any *linear* characteristic nor have they any *static* trait. The reaction of the different stakeholders to inputs that may come from the external world or from what happens inside the destination may be largely unpredictable as the outcomes of their conducts. Nonetheless, the system as a whole looks to follow some general "laws". Models such as the one by Butler (1980), although discussed, criticised, amended and modified (Hall & Butler, 1995; Lagiewski, 2005; McKercher, 2005; Russell, 2005), are generally considered able to give a meaningful description of a tourism destination and, in many cases, have proved useful tools for managing their development. Obviously, these are only limited considerations and the theoretical work in this field is still in its infancy. Just a handful of researchers have started to consider the complex systems approach as a more effective framework for the understanding of the many and different phenomena in this field (Farrell & Twining-Ward, 2004; Faulkner & Russell, 1997; McKercher, 1999). Much is still to be done, but the hopes are those of being able to understand, for example, how crises, disasters or turbulent changes may influence the sector, or why, after main crises, such as the 9/11 one, the tourism sector is able to show rapid and almost unexpected recoveries (Faulkner & Russell, 2001; Prideaux et al., 2003; Scott & Laws, 2005; UNWTO, 2002).

The McKercher's (1999) model quoted above, for example, looks very promising. It describes the main components and the operation of complex tourism systems with the aim of providing a representation of the elements that influence tourism on a wide range of possible scales: national, regional, local and, possibly, single enterprises. In fact, even if the number of actors influencing the system changes at each level, the relationships between the different elements are similar. This way, the author tends to provide a framework better able to explain, for example, the failure of many well designed, controlled and sustainable tourism development plans.

## 2.1 Complex systems evolution

A CAS is a dynamical system. It is, therefore, subject to some kind of evolution which may be characterised by two variables: an order parameter and a control parameter. The first one represents, in some way, the internal structure of the system, capturing its intrinsic order. The second one is an external variable which can be used to induce phase transitions in a system. For example, let us consider a certain volume of water close to the boiling point.

The order parameter is the density difference between the liquid and vapour phases; the temperature the control parameter. By increasing the temperature (providing energy, heath, to the system) it is possible to bring the water to the boiling point. At the critical temperature $T_c$ = 100 °C, the water starts boiling and the order parameter undergoes an abrupt change. It has the value zero in the random state (above the transition temperature) and takes on a nonzero value in the ordered state (below the transition).

More generally, the variation of the order parameter can lead the system to a critical point (bifurcation) beyond which several stable states may exist. The state will depend on small random fluctuations that are amplified by positive feedback. It is impossible to determine or to control which state will be attained in a specific empirical system; "*in practice, given the* *observable state of the system at the beginning of the process, the outcome is therefore unpredictable.*" (Heylighen, 2003: 12). Not even the control parameter (by itself) can be used to predict the system dynamics. Nonetheless, it is possible to sketch a general dependency of "global conditions" of a system on a control parameter.

Starting from a completely ordered and stable system, an increase in the control parameter will evolve it. The system passes through a periodic state, then to a situation characterised by a complex behaviour, then to a completely chaotic state. This last state can be adequately described with Wolf (1986: 273):

> In common usage chaos is taken to mean a state in which chance prevails. To the nonlinear dynamicist the word chaos has a more precise and rather different meaning. A chaotic system is one in which long-term [quantitative] prediction of the system's state is impossible because the omnipresent uncertainty in determining its initial state grows exponentially fast in time.

Most of the real systems we know live at the boundary between complexity and chaos. A situation frequently called *edge of chaos*, where a system is in a condition of fragile equilibrium, on the threshold of collapsing into a rapidly changing state, which may set off a new dynamic phase (Waldrop, 1992). The type of behaviour may depend on the initial state of the system and the values of its parameters, the boundaries are given by the critical values of the parameter. In the critical regions, called attractors, the system is locally stable. Overcoming a critical state we find a catastrophic bifurcation, then, as the evolution continues, the system moves towards a new attractor, waiting for the next perturbation able to create a bifurcation.

The history of a complex system is usually depicted by drawing its movement in the phase space. This is a geometrical n-dimensional space, in which the coordinates are the variables of the system. A dynamical system, at least in theory, can be described by a number of differential equations (equations of motion) comprising a number of variables. They are chosen in such a way that complete knowledge of all the variables determines the state of the system at one time in a unique way. The phase space is the set of all possible states of the system.

As time evolves, a point representing a system state in the phase space describes a trajectory (or orbit). The knowledge of this orbit implies the solution of the equations of motion. Stable orbits (attractors) mean stable system behaviours. This apparent continuity in the possible evolution of a system (from an orderly phase to a complex behaviour to a chaotic unpredictable dynamics) has led many to think of chaos and complexity phenomena as belonging to a "unified" discipline (Chris Langton quoted by Lewin 1999: 12):

> You are dealing with non-linear dynamical systems. In one case you may have a few things interacting, producing tremendously divergent behaviour. That's what you'd call deterministic chaos. It looks random, but it's not, because it's the result of equations you can specify, often quite simple equations. In another case interactions in a dynamical system give you an emergent global order, with a whole set of fascinating properties.

In other words: chaos theory essentially studies nonlinear effects on deterministic systems, while complexity theory studies definite patterns on non-deterministic systems. The focus of chaos theory is on the manner in which simple systems give rise to complicated unpredictable behaviours, while complexity theory focuses on how systems consisting of many elements can lead to well-organised and (almost) predictable behaviours.

## 2.2 The analysis of complex systems

The toolbox of the complexity scientist has today become quite crammed. Several techniques have been developed to deal with the task of describing a complex system. Many of them originate from the work of $19^{th}$ century scientists, but only modern computational facilities have made it possible to solve them. Following Amaral and Ottino (2004), we can group these tools in three main classes: nonlinear dynamics, statistical physics and network theory.

### 2.2.1 Nonlinear dynamics

A striking characteristic of complex systems is the nonlinearity of the interactions among the components. The main consequence is that the equations describing its behaviour (provided they exist) can be solved only in very rare cases. The work of Poincaré on the three body problem, at the end of the 19th century, had shown that even "simple" Newtonian systems involving more than two bodies may exhibit very complicated dynamics with almost unpredictable results arising from small variations of the initial conditions.

Since then, a number of mathematical techniques have been developed to approximate the solutions of the differential equations used to describe such systems. Only the availability of modern powerful computers, however, made it possible to find "solutions" (which, in nearly all cases, are obtained by numerical approximations). Much of the mathematics of chaos theory involves the repeated iteration of simple formulas, which would be impractical to do otherwise.

Nonlinear dynamic systems are capable of exhibiting self-organisation and *chaos*. This mechanism is called *deterministic chaos*, since the equations of motion which generate such erratic, and apparently unpredictable behaviour do not contain any random terms. Deterministic chaos refers to the irregular (chaotic) motion generated by a system whose evolution is governed by dynamic laws that uniquely determine the state of the system at all times from a knowledge of the system's previous history. The source of irregularity is the exponential divergence of initially close trajectories in a bounded region of phase-space. This divergence can be measured with the aid of the theory proposed by the 19$^{th}$ century Russian astronomer Aleksandr Mikhailovich Lyapunov (Kantz & Schreiber, 1997). In this sense, chaotic behaviour can be regarded as very complex dynamics.

This sensitivity to initial conditions is sometimes popularly called the *butterfly effect*, suggesting the idea that chaotic weather patterns can be altered by a butterfly flapping its wings. A practical implication is that it is essentially impossible to formulate long term predictions about the behaviour of a dynamic system: even if it would be possible to fix the initial conditions to a predetermined, finite accuracy, their errors would increase at an exponential rate. Examples of systems exhibiting nonlinear (chaotic) behaviour are: the atmosphere, the solar system, plate tectonics, turbulent fluids, mixing of coloured dyes, economies, stock markets, population growth or the "simple" double pendulum (Gleick, 1987; Waldrop, 1992).

### 2.2.2 Statistical physics

Statistical physics (or statistical mechanics) is one of the fundamental fields of physics. It uses statistical methods for addressing physical problems. A wide variety of issues, with an inherently stochastic nature, is treated in such a way. It provides a framework for relating the microscopic properties of individual atoms and molecules to the macroscopic ones of materials observed in every day life. Thermodynamics, and thermodynamic properties can be explained as a natural result of statistics and mechanics (classical and quantum). The main result, and power, of this approach is in the bypass of some classical mechanics problems, such as the impossibility of solving the three-body problem, by dealing with systems composed by a large number of elements, reasoning in terms of statistical ensembles. Moreover, it introduced the idea of discrete models and agent-based models (Wolfram, 2002).

In recent years, our understanding of phase transitions and critical phenomena has led to the development of two important new concepts: universality and scaling (Amaral & Ottino, 2004). Many physical systems exhibit universal properties that are independent of the specific form of the interactions among their constituents. This, for analogy, may suggest the hypothesis that universal laws or results may also show up in other types of complex systems: social, economic or biological. The scaling hypothesis, born in the framework of the study of critical phenomena, has provided the idea that a set of relations, called scaling laws, may help in relating the various critical-point exponents characterising the singular behaviour of an order parameter and of response functions. The predictions of the scaling hypothesis are supported by a wide range of experimental work, and also by numerous calculations on model systems.

The concept of universality in statistical physics and complex systems has the basic objective of capturing the essence of different systems and classifying them into distinct classes. The universality of critical behaviour pushes the investigations on the features of the microscopic relationships important for determining critical-point exponents and scaling functions. Statistical approaches can be very effective in systems when the number of degrees of freedom (and elements described by a number of variables) is so large that an exact solution is not practical or possible. Even in cases where it is possible to use analytical approximations, most current research utilises the processing power of modern computers to simulate numerical solutions.

One more important outcome of the use of statistical physics methods is the use of discrete models. The fundamental assumption is that some phenomena can be modelled in terms of computer programs (algorithms) rather than in terms of analytical expressions. Cellular automata (see for example: Mitchell et al., 1993; Wolfram, 2002) are an example of discrete time and space models developed for a computer utilisation. Cellular automata are dynamic structures, discrete in space and time. They operate on a uniform, regular lattice, characterised by "local" interactions and are made up of many cells, each of which may be in one of a finite number of states. A cell may change state only at fixed, regular intervals, and only in accordance with fixed rules that depend on their own values and the values of neighbours within a certain distance. Applications exist in many fields of physical, chemical, biological and social sciences; propagation of fire, predator-prey models or the evolution of artificial organisations can be represented with cellular automata (Mitchell et al., 1993; Wolfram, 2002).

### 2.2.3 Network theory

Most complex systems can be described as networks of interacting elements. In many cases these interactions lead to global behaviours that are not observable at the level of the single elements and that share the characteristics of emergence typical of a complex system. Moreover, the collective properties of dynamic systems composed of a large number of interconnected parts are strongly influenced by the topology of the connecting network. The mathematical models of network structures have been developed in graph theory. A graph is a generalisation of the concept of a set of dots (vertices, nodes), connected by links (edges, arcs). These, depending on the specific situation may or may not have a direction (the graph is directed or undirected). In a directed graph it is possible to track a route from some vertex to another, but not in the opposite direction. Links may be associated with numeric values (they may represent distances, costs, energies, information exchanges etc.) called weights. In the last few years, a number of researchers have shed light on some topological aspects of many kinds of social and natural networks, (the WWW, power grids, collaboration networks, networks of words, metabolic networks, economic agents).

The first mathematical model, which has been used for many years to describe several kinds of networks, is due to Erdős and Rényi (1959; 1960). Their model (the ER model) represents a network as a set of nodes connected, two at a time, with probability $p$. The distribution of the nodes degrees $k$ (the number of connections per node) follows a Poisson law with a peak $\langle k \rangle$ (the average degree of the network):

$$P(k) \approx \frac{\langle k \rangle^k}{k!} e^{-\langle k \rangle}$$

The node degrees distribution $P(k)$ and the average degree characterise the network and may be used as a distinctive attribute. Other quantities commonly used to describe a graph are: the clustering coefficient $C$, which measures how close the neighbourhood of each node is to a complete subgraph (part of the graph in which every node is connected to all the others, also called clique) and the average length $L$ of a path between any two vertices.

More recently, a number of investigations started with the works of Watts and Strogatz (1998) and Barabási and Albert (1999) has provided evidence that, in many cases, real world networks are quite different from ER graphs. Following these, many other works have been published (see the reviews: Albert & Barabási, 2002; Boccaletti et al., 2006; Dorogovtsev & Mendes, 2002; Newman, 2003b; Watts, 2004).

The distribution *P(k)* of the nodes degrees can be used to classify the networks into three broad classes (Amaral et al., 2000):

- *single-scale networks*: in which *P(k)* exhibits exponential or Gaussian tails. This class contains the random ER graphs and the small world (SW) networks described by Watts and Strogatz (1998). They are characterised by large clustering coefficients and short average path lengths. Their degree distribution still follows a poissonian law;

- *scale-free networks*: *P(k)* has a power-law distribution: $P(k) \sim k^{-\gamma}$. The distribution is largely uneven, there is no characteristic mean nodal degree (the mean of a poissonian ER distribution), but some (few) nodes act as very connected hubs, having a very large number of ties, while the majority of nodes have a small number of links. Scale free (SF) networks (Barabási & Albert, 1999) are dynamic networks. They grow with the addition of new nodes and new links following certain mechanisms; the most commonly cited is a preferential attachment in which a new node has a higher probability to attach to one of the most connected ones;

- *broad-scale networks*: for which the degree distribution has a mixed behaviour, a power law regime followed by a sharp cut-off (exponential or Gaussian decay) of the tail.

Both SF and SW networks, very common structures among the real world networks, show peculiar characteristics such as (Newman, 2003a):

- *robustness*: stability of the system to random removal (or failure) of randomly chosen elements; but also

- *fragility*: high sensitivity to targeted attacks to the most connected hubs;

- *low internal friction*: extent and speed of *disease* (viruses, but also messages, fads, beliefs etc.) transmission are greatly improved with respect to a random ER network, in some cases it is shown that there are no critical thresholds at all for these phenomena.

We may expect that the topology of the tourism actors network, like many other social and economic networks, exhibits structures like the ones discussed above.

# 3   Tourism destinations as complex systems, the symptoms

A tourism destination is a complex agglomeration of diverse systems of interrelated economic, social and environmental phenomena and networks. As mentioned in the introduction, some authors have proposed this interpretation as more effective in providing insights in the structure and the dynamical evolution of the system. The objective of this section is to provide, besides what is already qualitatively discussed in the literature, some quantitative evidence for these ideas. The most distinctive characteristics of a complex adaptive system will be discussed: self-organisation and self-similarity, robustness and resilience and *edge of chaos* behaviours.

## 3.1   Self-organisation and self-similarity

In economic and social contests, self-organisation is seen through the spontaneous formation of structures such as associations, consortia etc., that group stakeholders in order to better cope with the environment, namely to share experiences, resources and to support one another in facing possible adversities. The basic idea is that open systems will reorganise at critical points of instability almost independently from external "re-ordering" actions.

Modern tourism starts in the second half of the 19$^{th}$ century and its development is strongly influenced by two factors: the availability of free time and the development of faster and cheaper transportation means. In few years the phenomenon assumes an important dimension. As soon as the mass of tourists and travellers starts to be significant, the system of tourism operators, in some cases involving travellers and tourists as well, begins to form associations or consortia. All the modern developed tourism destinations have seen this process. In Italy, for example, the last decade of the 19$^{th}$ century sees the birth of the local tourism promotion associations, the *pro-loco*, immediately followed by the foundation of the Touring Club Italiano (1894), the Società Italiana degli Albergatori (hotel association, 1894), the Automobile Club Italiano (1898), the Associazione Nazionale per il Movimento dei Forestieri (association for the movement of foreigners, 1900). The process continues with the setup of the Ente Nazionale per l'Incremento delle Industrie Turistiche (1919), which will later become the Italian National Tourist Board (Paloscia, 1994).

More recently, the same phenomenon has been observed in the transition to the market by the countries formerly subjected to the authority of the Soviet regime (Recanatini & Ryterman, 2001). After the disruption of the USSR, a number of business associations emerges to mitigate the initial output decline. Following the ideas of complexity theories, this process can be interpreted as a rational response to coordinate activities in a newly decentralised economy to face the initial "disorganisation".

This self-organisation of a tourism destination's stakeholders gives a different foundation to the initial stages of the Butler's (1980) lifecycle model. According to this, in the first phases, as soon as the *tourism phenomenon* gains some momentum, we witness the involvement of the local community, the build-up of facilities and infrastructures and the creation of tourism associations. This behaviour is quite typical of a complex system. Studies (Helbing & Vicsek, 1999) have confirmed that, even in presence of repulsive exchanges, a system tends to minimise the rate and the intensity of interactions giving rise to the spontaneous formation of agglomerations. In other words, as individual entities are trying to maximise their *own* success, these systems tend to reach a state with the highest *global* success, which is not trivial at all. If a system self-organises at all, it is also (more or less) symmetric and, hence, behaves (almost) optimally.

From a more quantitative point of view, self-organising systems are characterised by self-similarity and fractal geometries, in which similar patterns are repeated with different sizes or time scales without changing their essential nature. A power law relationship in some statistical parameter is one of the common signatures of a nonlinear dynamic process, which is, at a point, self-organised (Komulainen, 2004). Zipf's law and Pareto law, for example, are well known principles, both exhibiting a power law behaviour and are commonly considered to be tell-tale signs of self-organisation.

Ulubaşoğlu and Hazari (2004) have shown the presence of Zipf-like relationships in tourism systems. In their work, the authors analyse the international tourist arrivals and find the familiar nonlinear rank-size distribution. With the same technique the data for tourist arrivals to Italian provinces have been analysed (ISTAT, 2005). The results are given in figure 1. The comparison of the two sets of data shows a remarkable similarity (apart from the obvious difference in scale).

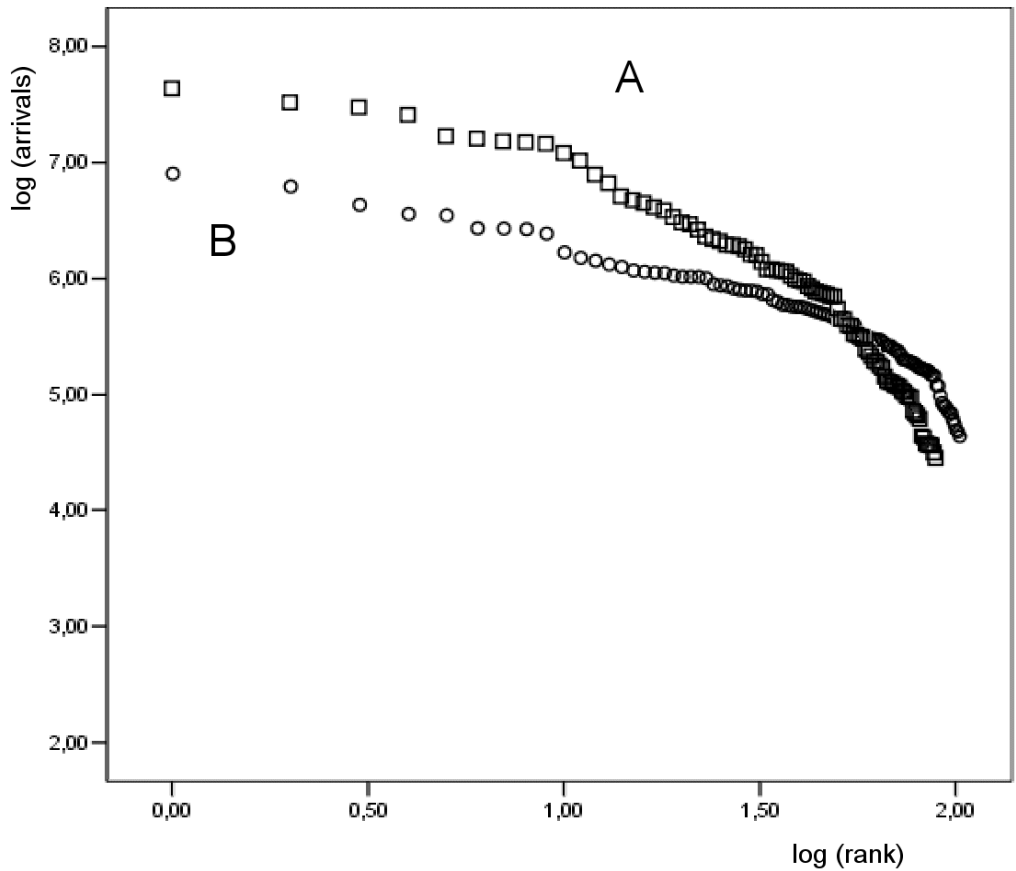

Figure 1: The Zipf-like distributions of tourist arrivals for World countries (marked A, after Ulubaşoğlu and Hazari, 2004) and Italian provinces (marked B, after ISTAT, 2005). Apart from the obvious scale difference, the overall behaviour is remarkably similar.

### 3.2 Robustness and resilience

The autonomous reorganisation capabilities of a complex system, exercised to react to external impulses that may disrupt it, reveal themselves as resilience. The concept, borrowed from the materials engineering field, has been applied by Holling (1973) in studying ecosystems. In this context, resilience represents the ability of a system to absorb disturbance and to reorganise while going through a change but retaining essentially the same functions, structure and characteristics. In a complex system, resilience is thus related to the magnitude of a shock that the system can absorb while remaining within a given state, or the to degree to which the system is capable of self-organisation. This is related to the extent to which reorganisation is endogenous rather than forced by external drivers. Resilience stems from both the internal structure of the system and the stimulus of private or public policy decisions (Mileti, 1999).

The dynamics of a system can be represented by measuring some outcome at regular intervals. In a tourism destination, the measurements of the days spent by tourists can be taken as a meaningful representation of the system. They may be seen as the result of the interactions among many of the system's components: the demand side (tourists), the supply side (infrastructures available to visitors) plus a number of internal and external economic factors (see for example: Ferro Luzzi & Flückiger, 2003).

An important concept in the analysis of a time series is the one of stationarity. From a statistical point of view its importance is due to the fact that many models used to study a time series and to forecast future behaviours are valid only in the presence of this characteristic (Chatfield, 1996). In our case we may extend this concept and give a physical interpretation. Stationarity can measure the capability of a system to continue its evolution absorbing possible external (or internal) shocks. This can be true for the whole series or for the parts that satisfy this requirement. In other words, if a tourism system, in its measurable expression, exhibits a consistent stationarity, at least for a reasonably long period of time, it means that, in that period of time, the system is able to recover disturbances in a relatively fast way. Possible deviations from this behaviour will be seen as structural breaks in the time series, with a sensible change in the series trend and/or level. In this case, we may conclude that the resilience of the system was not enough to react to shocks of the magnitude experienced.

Different techniques can be used to measure this kind of effect. Structural time series modelling (Harvey, 1989), for example, is used by Eugenio-Martin et al. (2005) to show that some part of the tourism demand in Scotland has been hardly affected by international crises. More than that, a number of proposals has been made to determine whether a time series can be considered exhibiting substantial stationarity. This set of statistical tests comprises the well known Dickey-Fuller test (Dickey & Fuller, 1979), both in the simple (DF) and the *augmented* (ADF) version, and the variations proposed by Phillips and Perron (PP test: Phillips & Perron, 1988) and by Zivot and Andrews (ZA test: Zivot & Andrews, 1992) and the more recent test by Lee and Strazicich, (LS test: Lee & Strazicich, 2003) based on Lagrange multipliers. Metes (2005) provides a good description of the tests and a comparison of their applicability, limitations and power.

The main idea behind these tests is that, after having allowed for a weak trend, the time series is checked against the hypothesis of modifications or breaks in the trend (or in the level) that can *shock* the system preventing it from returning to the previous behaviour and permanently moving it away from its past *track*. Several examples of these kind of studies can be found in the literature. Aly and Strazicich (2002) use the LS test to examine the annual tourist night visits to Egypt and Israel. They conclude that, in spite of shocks from terrorism, war and regional instability, visits by tourists remains a trend-reverting series. Narayan (2005) applies a modified ZA test proposed by Sen (2003) and shows that visitor arrivals in Fiji from Australia, New Zealand and the USA are trend-reverting at the 10% level or better, implying that shocks (due to internal political instabilities) to visitor arrivals have only a transitory effect.

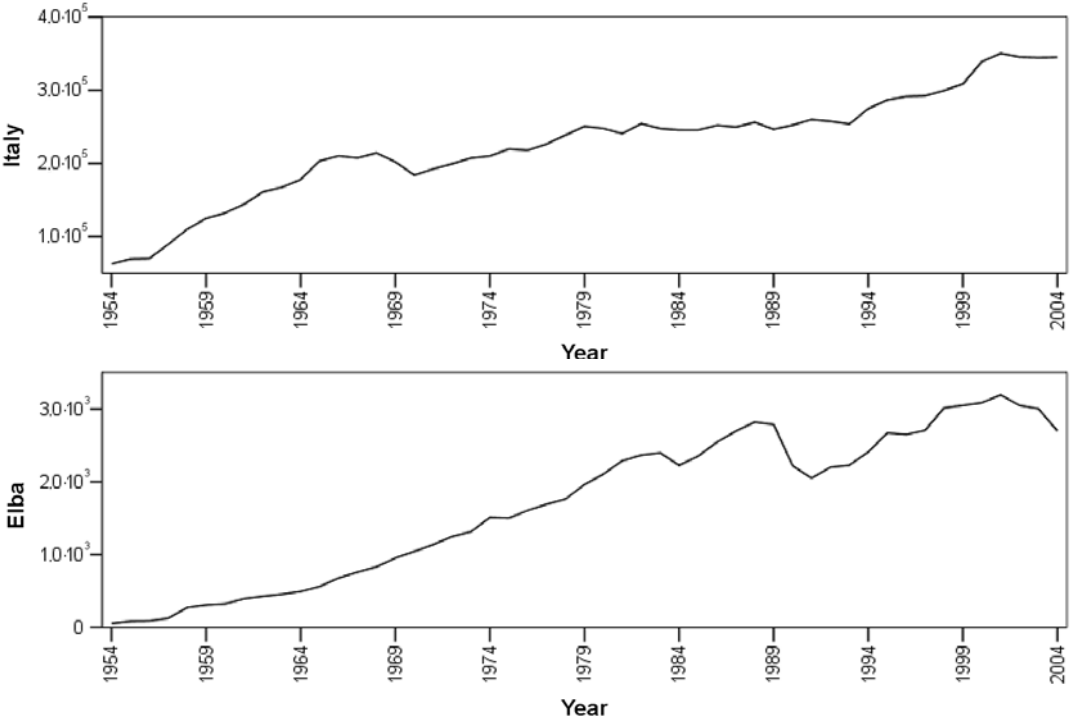

Figure 2: Tourist overnights time series for the period 1954-2004: Italy and Elba island (source: Italian Statistics Bureau, ISTAT and Elba Tourist Board)

Time series such as those shown in figure 2 (overnights for Italy and Elba Island, a renown Italian tourism destination), like many others we could draw for a tourism destination, are intrinsically not stationary, thus implying a dynamical evolution of the system. Removing the main trend and correcting seasonality effects, it is possible to assume that we are left with a quantity which represents the internal dynamics of the system's components. By applying the tests quoted above it is possible to analyse the response of the system. Tab. 1 shows the results for the two series.

Table 1: Stationarity tests results for Italy and Elba overnights time series. Data is for the period 1954-2004 [($^1$), ($^2$), ($^3$) indicate 10%, 5%, and 1% significance in the rejection of the unit root hypothesis].

| Destination | Test | | | |
|---|---|---|---|---|
| | ADF | PP | ZA | LS |
| Italy | -3.168 ($^2$) | -3.195 ($^2$) | -3.529 ($^2$) | -9.462 ($^3$) |
| Elba | -1.287 | -3.652 ($^3$) | -4.779 ($^3$) | -5.565 ($^1$) |

The results clearly suggest the rejection of the hypothesis of unit roots in the time series, thus confirming the basic stationarity of the systems' behaviours. The only conflicting result is the ADF test for Elba, but this test is known in literature as performing poorly in presence of possible structural breaks such as the ones that can be seen in the Elba series (Lee & Strazicich, 2003; Phillips & Perron, 1988; Zivot & Andrews, 1992). These results can therefore be interpreted as indication of a substantial resilience of the tourism systems analysed. Moreover, a system (Italy) and one of its subsystems (Elba) exhibit the same behaviour; an evident symptom of self-similarity.

## 3.3   On the edge of chaos

A tourism destination is an evolving, growing, system. Since the proposal by Butler (Butler, 1980) a fair amount of studies has given confirmations and examples of his model describing the life cycle of such systems (see for example: Butler, 2005a, 2005b). To preserve its existence, a CAS must interact with the environment, continually maintaining a flow of energy into and out of the system. Open systems that evolve and grow by absorbing energy and matter from the external world and dissipating the resulting entropy, are called *dissipative structures* (Prigogine & Nicolis, 1977). They evolve in unstable environments; the more ordered and complex a system becomes, the more entropy it must dissipate to maintain its existence. Each system has an upper limit on the amount of entropy which can be dissipated. When the flux from the environment increases beyond that limit, the system goes into a chaotic state. At a critical point, the system faces a bifurcation: it can break down, ceasing to exist as an organised system, or it can undergo a spontaneous reordering (a self-organisation). In a dissipative system, disequilibrium is a necessary condition for growth (Prigogine & Nicolis, 1977). On the other hand, managing a system typically aims at stabilising it trying to avoid abrupt changes. An organisation is therefore "oscillating" between a stable ordered state and a chaotic one, this complex state has been also called the *edge of chaos* (although the usage of this term is questioned by some, see for example: Mitchell et al., 1993).

It is possible to visualise this state by using a phase space plot, in which the behaviour of the system is rendered through the drawing of the parameters that characterise the system itself. A time series can be used to derive such a plot. Before doing that, we must recreate the phase space by using one of the techniques devised for this purpose. The most commonly used is the time-lagged (delay-coordinate) technique (Kantz & Schreiber, 1997; Schreiber, 1999). A delay coordinate reconstruction can be obtained by plotting the time series versus a time-delayed version of it. For a 2-dimensional reconstruction, it is possible to plot the delay vector $y_n = (t_n, t_{n-V})$, where V is the lag or sampling delay: the difference between the adjacent components of the delay vector measured in number of samples. The theoretical basis for this procedure is due to Takens (1980). His fundamental theorem states that a dynamical system can be reconstructed from a sequence of observations of the state of the dynamical system and, in the general case, the dynamic of the system recovered by "time-lagging" the series is the same as the dynamic of the original system.

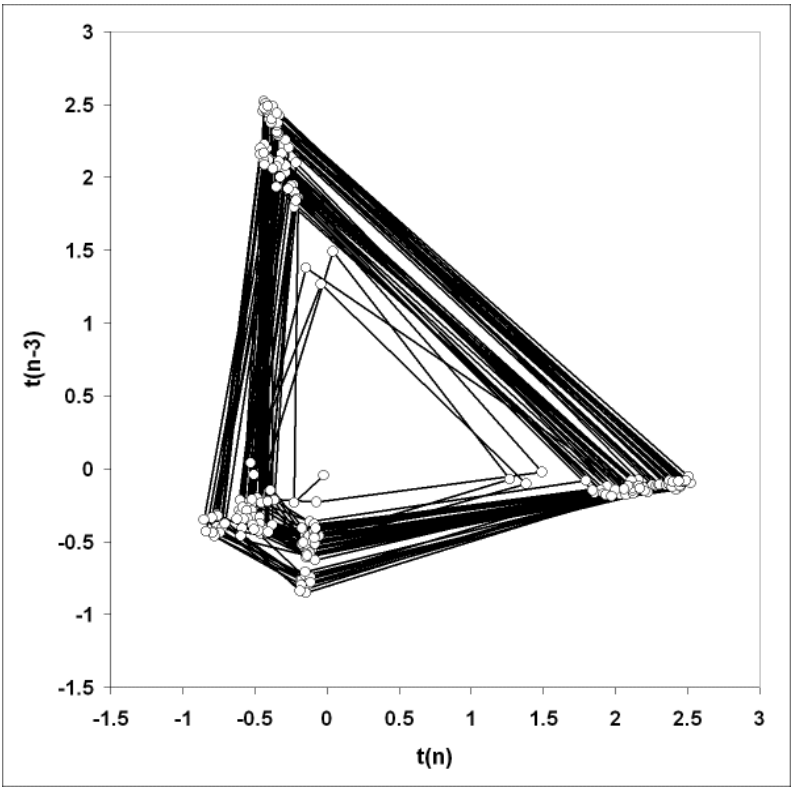

Figure 3: Phase space plot of quarterly tourist overnights for the period 1954-2004, Elba Island (data source: Elba Tourist Board)

As an example, the time series of quarterly tourist overnights at Elba Island for the period 1954-2005 has been considered. The series has been logarithmically transformed and "traditional" corrections for trend and seasonality have been applied. This way we may suppose that the pattern left is mainly (if not fully) dependent on the *internal dynamics* of the system. The resulting series is then embedded (with delay = 3) and the resulting phase space plot is shown in figure 3.

The diagram indicates that the series cycles regularly, but not fully predictably, staying within a bounded region of the phase space. This is typical of chaotic systems such as the weather or financial markets. The limits of the behavioural range shown in the phase space diagram are said to attract the behaviour of the system, and, due to its characteristics, it is called a strange attractor. The same basic techniques used here allow deriving also a quantitative measurement of *chaoticity*. This is given by the Lyapunov characteristic exponent (LCE).

The LCE of a dynamical system is a quantity that characterises the rate of separation of infinitesimally close trajectories in phase space. Quantitatively, two trajectories with initial separation $\delta X_0$ diverge according to: $\delta X(t) \approx e^{\lambda t} \delta X_0$. The rate of separation can be different for different orientations of the initial separation vector. Thus, there is whole spectrum of Lyapunov exponents. Their number is equal to the number of dimensions of the phase space. The largest LCE determines the general behaviour of the system. If it is negative, the system follows a stable trajectory, if it is null, the system is in a steady state, if it is positive the system exhibits unstable and chaotic behaviour (Sprott, 2003). In most cases, the calculation of Lyapunov exponents cannot be carried out analytically and numerical techniques must be used. In cases like ours, when only a 1-dimensional time series is given, the highest LCE can be estimated with the method proposed by Wolf et al. (1985) and Rosenstein et al. (1993).

The result for Elba time series is: LCE = 0.0192 (embedded with delay = 3 and dimension = 4). Being positive, we may assume that the series represents a chaotic system. Nonetheless, the value is quite small. As a comparison a logistic map has LCE = 0.69 and a Lorenz attractor has LCE = 0.91 (Sprott, 2003). It is therefore possible to conclude that the *chaoticity level* of our tourism system is present but not very high.

## 4  Implications for the management of a tourism destination

Far from being *definite proofs*, the results and the considerations presented so far have given evidence (both measurable and qualitative) for the presence of a number of characteristics that are typical symptoms of the complexity of a system. This approach to a tourism destination has its main outcome in explaining, at a conceptual level, a great part of the variability in the behaviour of these systems. In dealing with such a system, the "manager" may better understand the reasons of a limited, and sometimes inexistent, capacity of accurate prediction of future evolution which affects greatly the ability to manage and control it.

First of all, the techniques presented here can be used to assess the complexity and chaotic characteristics of a destination. Once evaluated, these can provide important indications for a destination management organisation. Significant issue, basis for most of the activities aiming at defining policies and plans for the development of a tourism destination, is the capability of forecasting, with the best possible approximation, its future behaviour and outcomes in terms of tourists' fluxes and receipts. One of the most important characteristics of a complex system, as said above, is its unpredictability, at least with the *linear* techniques we are used to.

Nonetheless, it is still possible to manage and understand complex systems, at least on some levels. Large scale behaviours at system level might still be foreseeable if it is possible to describe the overall dynamics of the system including the presence of any attractors and their basins (the regions of phase space in which they act). This can be accomplished by using a number of modelling approaches based on simulation models and nonlinear time series analysis. Once the attractors of a complex system have been identified, it can be possible to determine whether changes in a control parameter can produce sudden shifts in behaviour, or at least establish a probability distribution for their occurrence (Hansell et al., 1997). Simulation methods prove the feasibility of this approach (Smith & Johnson, 2004) and help in better capturing the complex behaviours of many systems composed of human organisations (Berry et al., 2002).

In recent years quite a number of these simulations has been used to describe different types of systems. Agent-based modelling (ABM) has proved its capabilities in a wide range of areas and has been proposed successfully for applications both in economic (Tesfatsion & Judd, 2006) and social sciences (Bonabeau, 2002; Macy & Willer, 2002). In ABM, a system is represented by a collection of autonomous decision-making units called agents. Each one independently estimates its situation and decides on the basis of a given set of rules. Agents may perform different tasks, depending on the system they represent. Repetitive interactions between agents produce different configurations of the system's state. Essential to the method is the extensive usage of computational capabilities. Agent-based models are especially appropriate when aggregate behaviour depends on structures of relations, so that no single representative element can fully describe the system under study.

The technique has been used as effective decision-making support for planning urban transportation and for analysing disease spread scenarios (Toroczkai & Eubank, 2005) or in studying travel demand patterns (Zhang & Levinson, 2004). More importantly, ABM has been effectively employed as a tool for providing the bases for the analysis and the development of policies in complex and uncertain socio-economic systems (Bankes, 1993, 2002). Tourism has seen only a very few attempts at using these techniques in simulating the behaviour of a tourism destination (Walker et al., 1998) or in supporting strategic management decisions (Buchta & Dolnicar, 2003). The discussion and the evidence provided above may be seen as a foundation to this way of reasoning, but definitely more work is needed from both a theoretical and a *practical* point of view.

It is possible to summarise what has been presented so far by saying that, once established what kind and what level of complexity a destination has, simulation methods can be used to provide medium or (relatively) long term evolution scenarios. The more traditional (linear) forecasting methods, with their limited validity, can give useful short-term predictions, as they have up to now. Their usefulness will mainly reside in contributing to the identification of main evolutionary paths and, especially, in applying small "corrections" to the system behaviour to try to steer it clear from undesired regimes.

One more implication of the complexity approach is the understanding that all the attempts to maintain stability may only work for a short period of time. Seeking stable equilibrium relationships is considered to be detrimental for the development of the system, since evolution and growth can only be possible in regions of the phase space at the boundary between order and chaos (Rosenhead, 1998):

> Rather than trying to consolidate stable equilibrium, the organisation should aim to position itself in a region of bounded instability, to seek the edge of chaos. The organisation should welcome disorder as a partner, use instability positively. In this way new possible futures for the organisation will emerge, arising out of the (controlled) ferment of ideas which it should try to provoke. Instead of a perfectly planned corporate death, the released creativity leads to an organisation which continuously re-invents itself. Members of an organisation in equilibrium with its environment are locked into stable work patterns and attitudes; far from equilibrium, behaviour can be changed more easily.

For a tourism destination, as well as for other types of organisations, it is possible to state (Stacey, 1993, 1996) that the systems do not only adapt to their environments, but help to create them and their success can come from contradiction as well as consistency. We have seen that long-term planning is almost impossible, therefore it is possible to conclude that success may stem from being part of a self-adapting process, rather than from an explicit 'vision' and revolutionary as well as incremental changes may form the basis to build organisational success. Managing a complex system requires, therefore, an adaptive attitude, more than a rigid deterministic, authoritarian style.

The proposal of using *adaptive management* to deal with a system derives from the work of 1970's ecologists (Holling, 1978). It calls for an *experimental* path to management. The method builds on the idea of exploring alternative possibilities, implementing one or more of them, monitoring the outcomes, testing the predictions and learning which one most effectively allows to meet the management objectives. The cycle then closes by using the results of the actions to improve knowledge and adjust subsequent management activities.

Since then, it has been adopted in different situations, including tourism systems, with encouraging results (Farrell & Twining-Ward, 2004). For example, Agostinho and Teixeira de Castro (2003) analyse a Brazilian experience and provide tangible data showing that an adaptive, self-organising, management system produce better performance with respect to more traditional schemes and Reed (1999) reports on the achievements obtained in collaborative planning of a Canadian destination. Even if the difficulties and the risks of this approach have been well highlighted (see for example: Caffyn & Jobbins, 2003), it looks to be one of the only reasonable alternative to successfully steer a contemporary tourism destination in achieving its goals and objectives.

## 5 Conclusive remarks

For decades, researchers and practitioners have approached tourism management and planning activities by using a reductionist approach and have thought that such a complex system could have been understood by looking at its principal components. This paper has provided some evidence to the idea that tourism, and its main representative, a tourism destination, is a complex adaptive system. Quantitative considerations have been presented to improve the so far mainly qualitative treatments in the literature on this subject.

The usage of complexity theory in the analysis of a tourism destination aims at giving better insights into the nature and the evolutionary behaviour of the system. This point of view has several implications for the ways in which a destination may be managed. It is argued that a shift in management attitude is needed and that dynamic and adaptive methods may be better suited to deal with such systems.

One of the main consequences of the complex nature of these organisations is an inability to predict accurately the environment and the effects on the evolution of a tourism destination. However, it is firm conviction of the author that the *linear* methods developed so far for forecasting, planning and managing retain their validity provided that the limitations of the methods are fully understood. The different perspective presented here can be helpful in contributing more elements for the analysis. A combination of traditional methods with dynamical numeric simulations is deemed able to deliver a number of "reasonable" future scenarios. They could offer more effective tools to define policies and lines of action aiming at the balanced development of a tourism destination.

The main limitations of this work are in the theoretical bases of complexity theories. Still a young and evolving discipline, it has not been able, yet, to produce universally accepted frame of reference and methods. The hope is that in a near future, some advancement may give better tools and conceptual constructions. Work is under way to meet this objective.

**Acknowledgments**

The author wishes to thank Valeria Tallinucci and Carla Catastini for their precious help in collecting the statistical data used in this work and Magda Antonioli Corigliano, Chris Cooper and Noel Scott for the valuable discussions and support during the preparation of this paper. The author wishes also to gratefully acknowledge the helpful comments of an anonymous reviewer.

Prideaux, B., Laws, E., & Faulkner, B. (2003). Events in Indonesia: exploring the limits to formal tourism trends forecasting methods in complex crisis situations. *Tourism Management, 24*, 475-487.

Prigogine, I., & Nicolis, G. (1977). *Self-Organization in Nonequilibrium Systems*. New York: Wiley.

Procaccia, I. (1988). Complex or just complicated? *Nature 333*, 498-499.

Recanatini, F., & Ryterman, R. (2001). *Disorganization or Self-Organization? The Emergence of Business Associations in a Transition Economy* (Policy Research Working Paper No. 2539): The World Bank.

Reed, M. G. (1999). Collaborative Tourism Planning as Adaptive Experiments in Emergent Tourism Settings. *Journal of Sustainable Tourism, 7*(3&4), 331-355.

Reynolds, C. (1987). Flocks, herds, and schools: a distributed behavioral model. *Computer Graphics, 21*, 25-34.

Ritchie, J. R. B., & Crouch, G. I. (2003). *The Competitive Destination: A Sustainable Tourism Perspective*. Oxon, UK: CABI Publishing.

Rosenhead, J. (1998). *Complexity Theory and Management Practice. Science as Culture*. Retrieved December 2005, from http://www.human-nature.com/science-as-culture/rosenhead.html.

Rosenstein, M. T., Collins, J. J., & De Luca, C. J. (1993). A practical method for calculating largest Lyapunov exponents from small data sets. *Physica D, 65*, 117-134.

Russell, R. (2005). Chaos theory and its application to the Tourism Area Life Cycle Model. In R. W. Butler (Ed.), *The Tourism Area Life Cycle, Vol. 2: Conceptual and Theoretical Issues* (pp. 164-180). Clevedon, UK: Channel View.

Russell, R. (2006). Chaos theory and managerial approaches. In D. Buhalis & C. Costa (Eds.), *Tourism Dynamics, Challenges and Tools: Present and Future Issues* (pp. 108-115). Oxford Butterworth-Heinemann.

Russell, R., & Faulkner, B. (1998). Reliving the destination life cycle in Coolangatta: An historical perspective on the rise, decline and rejuvenation of an Australian seaside resort. In E. Laws, B. Faulkner & G. Moscardo (Eds.), *Embracing and Managing Change in Tourism: International Case Studies* (pp. 95-115). London: Routledge.

Russell, R., & Faulkner, B. (1999). Movers and Shakers: chaos makers in tourism development. *Tourism Management, 20*(4), 411-423.

Russell, R., & Faulkner, B. (2004). Entrepreneurship, Chaos and the Tourism Area Lifecycle. *Annals of Tourism Research, 31*(3), 556-579.

Saari, D. G. (1995). Mathematical Complexity of Simple Economics. *Notices of the American Mathematical Society, 42*(2), 222-230.

Schreiber, T. (1999). Interdisciplinary application of nonlinear time series methods. *Physics Reports, 308*, 1-64.

Scott, N., & Laws, E. (2005). Tourism Crises and Disasters: Enhancing Understanding of System Effects. *Journal of Travel & Tourism Marketing, 19*(2/3), 149-158.

Sen, A. (2003). On unit-root tests when the alternative is a trend-break stationary process. *Journal of Business and Economics Statistics, 21*, 174-184.

Smith, D. M. D., & Johnson, N. F. (2004). *Evolution Management in a Complex Adaptive System: Engineering the Future* (arXiv/cond-mat/0409036). Retrieved January, 2006, from http://arxiv.org/abs/cond-mat/0409036.

Smith, S. L. J. (1988). Defining Tourism, A Supply-Side View. *Annals of Tourism Research, 15*, 179-190.
31